\documentclass[12pt]{article}
\begin{document}
\title{
\vspace*{-30pt}
 CPT and effective Hamiltonians \\ for neutral kaon
and similar complexes\footnote{Talk given at the Conference on
{\em "Irreversible Quantum Dynamics"}, 29 July --- 2 August 2002,
Abdus Salam International Centre for Theoretical Physics, Trieste,
Italy. }}
\author{ \hfill \\ Krzysztof URBANOWSKI\footnote{
e--mail:
K.Urbanowski@proton.if.uz.zgora.pl;
K.Urbanowski@if.uz.zgora.pl}
\\  \hfill  \\
University of Zielona G\'{o}ra, Institute of Physics, \\
ul. Podg\'{o}rna 50, 65-246 Zielona G\'{o}ra, Poland.\date{July
31, 2002}} \maketitle
\begin{abstract}
We  begin with a discussion of  the general form and general CP--
and CPT-- transformation properties of the Lee--Oehme--Yang (LOY)
effective Hamiltonian for the neutral  kaon complex. Next, the
properties of the exact effective Hamiltonian for  this complex
are discussed. Using the Khalfin Theorem we show that the diagonal
matrix elements of the effective Hamiltonian governing the time
evolution in the subspace of states of an unstable particle and
its antiparticle need not be equal at for $t
> t_{0}$ ($t_{0}$ is the instant of creation of the pair) when the
total system under consideration is CPT invariant but CP
noninvariant. The unusual consequence of this result is that,
contrary to the properties of stable particles, the masses of the
unstable particle "1" and its antiparticle "2" need not be equal
for $t \gg t_{0}$ in the case of preserved CPT and violated CP
symmetries. We also show that there exists an approximation which
is more accurate than the LOY, and which leads to an effective
Hamiltonian whose diagonal matrix elements posses properties
consistent with the conclusions for the exact effective
Hamiltonian described above.
\end{abstract}\pagebreak[4]

\section{Introduction.}

The problem of testing CPT--invariance experimentally has
attracted the attention of physicist, practically since the
discovery of antiparticles. CPT symmetry is a fundamental theorem
of axiomatic quantum field theory which follows from locality,
Lorentz invariance, and unitarity \cite{cpt}. Many tests of
CPT--invariance consist in searching for decay process of neutral
kaons. All known CP--  and hypothetically  possible CPT--violation
effects in the neutral kaon complex are  described by solving the
Schr\"{o}dinger--like evolution equation \cite{Lee1} ---
\cite{improved} (we use $\hbar = c = 1$ units)
\begin{equation}
i \frac{\partial}{\partial t} |\psi ; t >_{\parallel} =
H_{\parallel} |\psi ; t >_{\parallel} \label{l1}
\end{equation}
for $|\psi ; t >_{\parallel}$ belonging to the subspace ${\cal
H}_{\parallel} \subset {\cal H}$ (where ${\cal H}$ is the state
space of the physical system under investigation), e.g., spanned
by orthonormal neutral  kaons states $|K_{0}>, \;
|{\overline{K}}_{0}>$, and so on, (then states corresponding to
the decay products belong to ${\cal H} \ominus {\cal
H}_{\parallel} \stackrel{\rm def}{=} {\cal H}_{\perp}$), and
nonhermitian effective Hamiltonian $H_{\parallel}$ obtained
usually by means of the  Lee-Oehme--Yang (LOY) approach (within
the Weisskopf--Wigner approximation (WW)) \cite{Lee1} ---
\cite{Comins}, \cite{improved}:
\begin{equation}
H_{\parallel} \equiv M - \frac{i}{2} \Gamma, \label{new1}
\end{equation}
where
\begin{equation}
M = M^{+}, \; \; \Gamma = {\Gamma}^{+}, \label{new1a}
\end{equation}
are $(2 \times 2)$ matrices.

The solutions of Eq. (\ref{l1}) can be written in matrix form and
such  a  matrix  defines  the evolution    operator (which is
usually     nonunitary) $U_{\parallel}(t)$ acting in ${\cal
H}_{\parallel}$:
\begin{equation}
|\psi ; t >_{\parallel} = U_{\parallel}(t) |\psi ;t_{0} = 0
>_{\parallel} \stackrel{\rm def}{=} U_{\parallel}(t) |\psi
>_{\parallel}, \label{l1a}
\end{equation}
where,
\begin{equation}
|\psi >_{\parallel} \equiv q_{1}|{\bf 1}> + q_{2}|{\bf 2}>,
\label{l1b}
\end{equation}
and $|{\bf 1}>$ stands for  the  vectors of the   $|K_{0}>,  \;
|B_{0}>$ type and $|{\bf 2}>$ denotes antiparticles  of particle
"1": $|{\overline{K}}_{0}>, \; |{\overline{B}}_{0}>$, $<{\bf
j}|{\bf k}> = {\delta}_{jk}$, $j,k =1,2$.

In many papers it is assumed that the real parts, $\Re (.)$, of
the diagonal matrix elements of $H_{\parallel}$:
\begin{equation}
\Re \, (h_{jj} ) \equiv M_{jj}, \; \;(j =1,2), \label{m-jj}
\end{equation}
where
\begin{equation}
h_{jk}  =  <{\bf j}|H_{\parallel}|{\bf k}>, \; (j,k=1,2),
\label{h-jk}
\end{equation}
correspond to the masses  of particle "1" and its antiparticle "2"
respectively \cite{Lee1} --- \cite{improved}, (and such an
interpretation of $\Re \, (h_{11})$ and $\Re \, (h_{22})$ will be
used in this paper), whereas the imaginary parts, $\Im (.)$,
\begin{equation}
-2 \Im \, (h_{jj}) \equiv {\Gamma}_{jj}, \; \;(j =1,2),
\label{g-jj}
\end{equation}
are interpreted as the decay widths of these particles \cite{Lee1}
--- \cite{improved}. Such an interpretation seems to be consistent
with the recent and the early experimental data for the neutral
kaon and similar complexes \cite{data}.

Relations between matrix elements of  $H_{\parallel}$  implied  by
CPT--transformation properties of the Hamiltonian $H$ of the total
system, containing  neutral  kaon  complex  as  a  subsystem,  are
crucial to designing CPT--invariance and CP--violation tests and
to proper interpretation of their results. The aim of this paper
is to examine the properties of the exact $H_{\parallel}$
generated by the CPT--symmetry of the total system under
consideration and independent of the approximation used and to
compare these properties of the exact and of the approximate
$H_{||}$.

\section{$H_{LOY}$ and CPT--symmetry.}

Now, let us consider briefly some properties of the LOY model. Let
$H$ be the total (selfadjoint) Hamiltonian, acting in $\cal H$
--- then  the total unitary evolution  operator $U(t)$ fulfills
the Schr\"{o}dinger equation
\begin{equation}
i \frac{\partial}{\partial t} U(t)|\phi > = H U(t)|\phi >,  \; \;
U(0) = I, \label{Schrod}
\end{equation}
where $I$ is the unit operator in $\cal H$, $|\phi > \equiv |\phi
; t_{0} = 0> \in {\cal H}$  is  the  initial  state  of  the
system:
\begin{equation}
|\phi  >  \equiv  |\psi  >_{\parallel}  \label{l2a}
\end{equation}
in  our case  $|\phi ;t> = U(t) |\phi >$. Let $P$ denote the
projection operator onto the subspace ${\cal H}_{\parallel}$:
\begin{equation}
P{\cal H} = {\cal H}_{\parallel}, \; \; \; P = P^{2} = P^{+},
\label{new2}
\end{equation}
then the subspace of decay products ${\cal H}_{\perp}$ equals
\begin{equation}
{\cal H}_{\perp}  = (I - P) {\cal H} \stackrel{\rm def}{=} Q {\cal
H}, \; \; \; Q \equiv I - P. \label{l7c}
\end{equation}
For the  case of neutral  kaons  or  neutral  $B$--mesons,  etc.,
the projector $P$ can be chosen as follows:
\begin{equation}
P \equiv |{\bf 1}><{\bf 1}| + |{\bf 2}><{\bf 2}|, \label{P}
\end{equation}
and the definition of $|K_{0}>$ and $|{\overline{K}}_{0}>$ is
analogous to the one used in the LOY theory for corresponding
vectors. In the LOY approach it is assumed that vectors $|{\bf
1}>$, $|{\bf 2}>$ considered above are eigenstates of $H^{(0)}$
for a 2--fold degenerate eigenvalue $m_{0}$:
\begin{equation}
H^{(0)} |{\bf j} > = m_{0} |{\bf j }>, \; \;  j = 1,2, \label{b1}
\end{equation}
where $H^{(0)}$ is the so called free Hamiltonian, $H^{(0)} \equiv
H_{strong} = H - H_{W}$, and $H_{W}$ denotes weak and other
interactions which are responsible for transitions between  the
eigenvectors of $H^{(0)}$, i.e., for the decay process. This means
that
\begin{equation}
[P, H^{(0)}] = 0. \label{new3}
\end{equation}

The condition guaranteeing the occurrence of transitions  between
subspaces ${\cal H}_{\parallel}$ and ${\cal H}_{\perp}$, i.e., the
decay process of states in ${\cal H}_{\parallel}$, can  be written
as follows
\begin{equation}
[P,H_{W}] \neq 0 , \label{r32}
\end{equation}
that is
\begin{equation}
[P,H] \neq 0 . \label{[P,H]}
\end{equation}
Usually, in LOY and related approaches, it is assumed that
\begin{equation}
{\Theta}H^{(0)}{\Theta}^{-1} = {H^{(0)}}^{+} \equiv H^{(0)} ,
\label{r31}
\end{equation}
where $\Theta$ is the antiunitary operator:
\begin{equation}
\Theta \stackrel{\rm def}{=} {\cal C}{\cal P}{\cal T}.
\label{new4}
\end{equation}
The subspace of neutral kaons ${\cal H}_{\parallel}$ is assumed to
be invariant under $\Theta$:
\begin{equation}
{\Theta} P {\Theta}^{-1} = P^{+} \equiv P. \label{9aa}
\end{equation}

In the kaon rest frame, the time evolution is governed by the
Schr\"{o}dinger equation (\ref{Schrod}), where the initial state
of the system has the form (\ref{l2a}), (\ref{l1b}). Within
assumptions (\ref{b1}) --- (\ref{r32}) the  Weisskopf--Wigner
approach, which is the source of the LOY method, leads to the
following formula for $H_{LOY}$ (e.g., see
\cite{Lee1,Lee2,Cronin,improved}):
\begin{eqnarray}
H_{LOY} = m_{0} P  - \Sigma (m_{0})
& \equiv & PHP - \Sigma (m_{0}), \label{b3} \\
& = & M_{LOY} - \frac{i}{2}{\Gamma}_{LOY} \label{b3a}
\end{eqnarray}
where it has been assumed that $<{\bf 1}|H_{W}|{\bf 2}> = <{\bf
1}|H_{W}|{\bf 2}>^{\ast} = 0$ (see \cite{Lee1} ---
\cite{improved}),
\begin{equation}
\Sigma ( \epsilon ) = PHQ \frac{1}{QHQ - \epsilon - i 0} QHP.
\label{r24}
\end{equation}
The matrix elements $h_{jk}^{LOY}$  of $H_{LOY}$ are
\begin{eqnarray}
h_{jk}^{LOY} & = &  H_{jk} - {\Sigma}_{jk} (m_{0} ) , \; \; \;
(j,k = 1,2) ,  \label{b5} \\
& = & M_{jk}^{LOY} - \frac{i}{2} {\Gamma}_{jk}^{LOY} \label{b5a}
\end{eqnarray}
where, in this case,
\begin{equation}
H_{jk} = <{\bf j} |H| {\bf k} > \equiv <{\bf j} |(H^{(0)} + H_{W}
)| {\bf k} > \equiv m_{0} {\delta}_{jk} + <{\bf j}|H_{W}|{\bf k}>
, \label{b6}
\end{equation}
and ${\Sigma}_{jk} ( \epsilon ) = < {\bf j} \mid \Sigma ( \epsilon
) \mid {\bf k} >$.

Now, if ${\Theta}H_{W}{\Theta}^{-1} = H_{W}^{+} \equiv H_{W}$,
that is if
\begin{equation}
[ \Theta , H] = 0, \label{[CPT,H]=0}
\end{equation}
then using, e.g., the following phase convention \cite{Lee2}
--- \cite{improved}
\begin{equation}
\Theta |{\bf 1}> \stackrel{\rm def}{=} - |{\bf 2}>, \;\;
\Theta|{\bf 2}> \stackrel{\rm def}{=} - |{\bf 1}>, \label{cpt1}
\end{equation}
and taking into account that $< \psi | \varphi > =
<{\Theta}{\varphi}|{\Theta}{\psi}>$, one easily finds from
(\ref{b3}) -- (\ref{b6})  that
\begin{equation}
{h_{11}^{LOY}}^{\Theta} - {h_{22}^{LOY}}^{\Theta} = 0,  \label{b8}
\end{equation}
and thus
\begin{equation}
M_{11}^{LOY} = M_{22}^{LOY}, \label{LOY-m=m}
\end{equation}
(where ${h_{jk}^{LOY}}^{\Theta}$ denotes the matrix elements of
$H_{LOY}^{\Theta}$ --- of the LOY effective Hamiltonian when the
relation (\ref{[CPT,H]=0}) holds), in the CPT--invariant system.
This  is  the standard result of the LOY approach and this is the
picture  which one meets in the literature \cite{Lee1}  ---
\cite{dafne}.

If it is assumed that the CPT--symmetry is not conserved in  the
physical system under consideration, i.e., that
\begin{equation}
[ \Theta , H] \neq 0, \label{e67}
\end{equation}
then    $h_{11}^{LOY}    \neq    h_{22}^{LOY}$.

It  is  convenient to  express the difference between
$H_{LOY}^{\Theta}$   and  the effective Hamiltonian $H_{LOY}$
appearing  within  the  LOY approach in the case  of nonconserved
CPT--symmetry as follows
\begin{eqnarray}
H_{LOY} & \equiv & H_{LOY}^{\Theta} + \delta H_{LOY}  \label{e68} \\
& = & \left(
\begin{array}{cc}
( M_{0} + \frac{1}{2} \delta M) - \frac{i}{2} ( {\Gamma}_{0} +
\frac{1}{2} \delta \Gamma ),
& M_{12} - \frac{i}{2} {\Gamma}_{12} \\
M_{12}^{\ast} - \frac{i}{2} {\Gamma}_{12}^{\ast} & (M_{0} -
\frac{1}{2} \delta M) - \frac{i}{2} ({\Gamma}_{0} - \frac{1}{2}
\delta \Gamma )
\end{array}
\right) .  \nonumber
\end{eqnarray}
In other words
\begin{equation}
h_{jk}^{LOY} = {h_{jk}^{LOY }}^{  \Theta} + \Delta h_{jk}^{LOY},
\label{LOY=h+delta}
\end{equation}
where
\begin{equation}
\Delta h_{jk}^{LOY} = (-1)^{j+1}\frac{1}{2}(\delta M - \frac{i}{2}
\delta \Gamma ) {\delta}_{jk}, \label{LOY-delta}
\end{equation}
and $j,k = 1,2$. Within this approach the $\delta M$ and $\delta
\Gamma$ terms violate the CPT--symmetry.

\section{CPT and the exact effective Hamiltonian}

The aim of this Section is to show that, contrary to the LOY
conclusion (\ref{b8}), the diagonal matrix elements of the exact
effective Hamiltonian $H_{||}$ can not be equal when the total
system under consideration is CPT invariant but CP noninvariant.
This will be done by means of the method used in \cite{plb2002}.

The universal properties of the (unstable) particle--antiparticle
subsystem of the system described by the Hamiltonian $H$, for
which  the relation (\ref{[CPT,H]=0}) holds, can be extracted from
the matrix elements of the exact $U_{||}(t)$ appearing in
(\ref{l1a}). Such $U_{||}(t)$ has the following form
\begin{equation}
U_{||}(t) = P U(t)P, \label{U||}
\end{equation}
where $P$ is defined by the relation (\ref{P}), and $U(t)$ is the
total unitary evolution operator $U(t)$, which solves the
Schr\"{o}\-din\-ger equation (\ref{Schrod}). Of course,
$U_{||}(t)$ has a nontrivial form only if (\ref{[P,H]}) holds, and
only then transitions of states from ${\cal H}_{||}$ into ${\cal
H}_{\perp}$ and vice versa, i.e., decay and regeneration
processes, are allowed.

Using the matrix representation one finds
\begin{equation}
U_{||}(t) \equiv \left(
\begin{array}{cc}
{\rm \bf A}(t) & {\rm \bf 0} \\
{\rm \bf 0} & {\rm \bf 0}
\end{array} \right)
\label{A(t)}
\end{equation}
where ${\rm \bf 0}$ denotes the suitable zero submatrices and a
submatrix ${\rm \bf A}(t)$ is the $2 \times 2$ matrix acting in
${\cal H}_{||}$
\begin{equation}
{\rm \bf A}(t) = \left(
\begin{array}{cc}
A_{11}(t) & A_{12}(t) \\
A_{21}(t) & A_{22}(t)
\end{array} \right) \label{A(t)=}
\end{equation}
and $A_{jk}(t) = <{\bf j}|U_{||}(t)|{\bf k}> \equiv <{\bf
j}|U(t)|{\bf k}>$, $(j,k =1,2)$.

Now, assuming (\ref{[CPT,H]=0}) and using the phase convention
(\ref{cpt1}), \cite{Lee1} --- \cite{Comins}, one easily finds that
\cite{chiu}, \cite{nowakowski,leonid1,leonid2}
\begin{equation}
A_{11}(t) = A_{22}(t). \label{A11=A22}
\end{equation}
Note that assumptions (\ref{[CPT,H]=0}) and (\ref{cpt1}) give no
relations between $A_{12}(t)$ and $A_{21}(t)$.

The important relation between amplitudes $A_{12}(t)$ and
$A_{21}(t)$ follows from the famous Khalfin's Theorem \cite{chiu},
\cite{leonid1} --- \cite{leonid2}. This Theorem states that in the
case of unstable states, if amplitudes $A_{12}(t)$ and $A_{21}(t)$
have the same time dependence
\begin{equation}
r(t) \stackrel{\rm def}{=} \frac{A_{12}(t)}{A_{21}(t)} = {\rm
const} \equiv r, \label{r=const},
\end{equation}
there must be $|r| = 1$.

For unstable particles  relation (\ref{A11=A22}) means that the
decay laws
\begin{equation}
p_{j}(t) \stackrel{\rm def}{=} |A_{jj}(t)|^{2}, \label{p-j}
\end{equation}
(where $j = 1,2$), of the particle $|{\bf 1}>$ and its
antiparticle $|{\bf 2}>$ are equal,
\begin{equation}
p_{1}(t) \equiv p_{2}(t). \label{p1=p2}
\end{equation}
The consequence of this last  property is that the decay rates of
the particle $|{\bf 1}>$ and its antiparticle $|{\bf 2}>$ must be
equal too.

From (\ref{A11=A22}) it does not follow that the masses of
particle "1" and the antiparticle "2" should be equal.

More conclusions about the properties of the matrix elements of
$H_{||}$ one can infer analyzing the following identity
\cite{horwitz} --- \cite{pra}
\begin{equation}
H_{||} \equiv H_{||}(t) = i \frac{\partial U_{||}(t)}{\partial t}
[U_{||}(t)]^{-1}, \label{H||2a}
\end{equation}
where $[U_{||}(t)]^{-1}$ is defined as follows
\begin{equation}
U_{||}(t) \, [U_{||}(t)]^{-1} = [U_{||}(t)]^{-1} \, U_{||}(t) \, =
\, P. \label{U^-1}
\end{equation}
(Note that the identity (\ref{H||2a}) holds, independent of
whether $[P,H] \neq 0$ or $[P,H]=0$). The expression (\ref{H||2a})
can be rewritten using the matrix ${\bf A}(t)$

\begin{equation}
H_{||}(t) \equiv  i \frac{\partial {\bf A}(t)}{\partial t} [{\bf
A}(t)]^{-1}. \label{H||2b}
\end{equation}
Relations (\ref{H||2a}), (\ref{H||2b}) must be fulfilled by the
exact as well as by every approximate effective Hamiltonian
governing the time evolution in every two dimensional subspace
${\cal H}_{||}$ of states $\cal H$ \cite{horwitz} --- \cite{pra}.

It is easy to find from (\ref{H||2b}) the general formulae for the
diagonal matrix elements, $h_{jj}$, of $H_{||}(t)$, in which we
are interested. We have
\begin{eqnarray}
h_{11}(t) &=& \frac{i}{\det {\bf A}(t)} \Big( \frac{\partial
A_{11}(t)}{\partial t} A_{22}(t) - \frac{\partial
A_{12}(t)}{\partial t} A_{21}(t) \Big), \label{h11=} \\
h_{22}(t) & = & \frac{i}{\det {\bf A}(t)} \Big( - \frac{\partial
A_{21}(t)}{\partial t} A_{12}(t) + \frac{\partial
A_{22}(t)}{\partial t} A_{11}(t) \Big). \label{h22=}
\end{eqnarray}
Now, assuming (\ref{[CPT,H]=0}) and using the consequence
(\ref{A11=A22}) of this assumption, one finds
\begin{equation}
h_{11}(t) - h_{22}(t) =  \frac{i}{\det {\bf A}(t)} \Big(
\frac{\partial A_{21}(t)}{\partial t} A_{12}(t) - \frac{\partial
A_{12}(t)}{\partial t} A_{21}(t) \Big). \label{h11-h22=}
\end{equation}
Next, after some algebra one obtains
\begin{equation}
h_{11}(t) - h_{22}(t) = - i \, \frac{A_{12}(t) \, A_{21}(t) }{\det
{\bf A}(t)} \; \frac{\partial}{\partial t} \ln
\Big(\frac{A_{12}(t)}{A_{21}(t)} \Big). \label{h11-h22=1}
\end{equation}
This result means that in the considered case for $t>0$ the
following Theorem holds:
\begin{equation}
h_{11}(t) - h_{22}(t) = 0 \; \; \Leftrightarrow \; \;
\frac{A_{12}(t)}{A_{21}(t)}\;\; = \; \; {\rm const.}, \; \; (t >
0). \label{h11-h22=0}
\end{equation}
Thus for $t > 0$ the problem under study is reduced to the
Khalfin's Theorem (see the relation (\ref{r=const})).

From (\ref{h11=}) and (\ref{h22=}) it is easy to see that at $t=0$
\begin{equation}
h_{jj}(0) = <{\bf j}|H|{\bf j}>, \; \; (j=1,2), \label{hjjt=0}
\end{equation}
which means that in a CPT invariant system (\ref{[CPT,H]=0}) in
the case of pairs of unstable particles, for which transformations
of type (\ref{cpt1}) hold
\begin{equation}
M_{11}(0) = M_{22}(0) \equiv <{\bf 1}|H|{\bf 1}>, \label{M11=M22}
\end{equation}
the unstable particles "1" and "2" are created at $t=t_{0} \equiv
0$ as  particles with equal masses.

Now let us go on to analyze the  conclusions following from the
Khalfin's Theorem. CP noninvariance requires that $|r| \neq 1$
\cite{chiu,nowakowski,leonid1,leonid2} (see also \cite{Lee1}
--- \cite{Cronin}, \cite{data}). This means that in such a case
there must be $r \equiv r(t) \neq {\rm const.}$. So, if in the
system considered the property (\ref{[CPT,H]=0}) holds but
\begin{equation}
[{\cal CP}, H] \neq 0, \label{[CP,H]}
\end{equation}
and the unstable states "1" and "2" are connected by a relation of
type (\ref{cpt1}), then at $t > 0$ it must be $(h_{11}(t) -
h_{22}(t)) \neq 0$ in this system. Assuming the LOY interpretation
of $\Re \,(h_{jj}(t))$, ($j=1,2$), one can conclude from the
Khalfin's Theorem and from the property (\ref{h11-h22=0}) that if
$A_{12}(t), A_{21}(t) \neq 0$ for $t > 0$ and if the total system
considered is CPT--invariant, but CP--noninvariant, then
$M_{11}(t) \neq M_{22}(t)$ for $t >0$, that is, that contrary to
the case of stable particles (the bound states), the masses of the
simultaneously created unstable particle "1" and its antiparticle
"2", which are connected by the relation (\ref{cpt1}), need not be
equal  for $t
>t_{0} =0$.  Of course, such a conclusion contradicts
the standard LOY result (\ref{b8}), (\ref{LOY-m=m}). However, one
should remember that the LOY description of neutral $K$ mesons and
similar complexes is only an approximate one, and that the LOY
approximation is not perfect. On the other hand the relation
(\ref{h11-h22=0}) and the Khalfin's Theorem follow from the basic
principles of the quantum theory and are rigorous. Consequently,
their implications should also be considered rigorous.

\section{Beyond the LOY approximation}

The approximate formulae for $H_{\parallel}(t)$ have been  derived
in \cite{9,10}  using  the  Krolikowski--Rzewuski equation   for
the projection of a state vector \cite{7}, which results from the
Schr\"{o}dinger  equation (\ref{Schrod}) for  the  total system
under consideration, and, in the  case  of the initial conditions
of the type (\ref{l2a}), takes the following form
\begin{equation}
( i \frac{\partial}{ {\partial} t} - PHP ) U_{\parallel}(t)|\psi
>_{||}
 =  - i \int_{0}^{\infty} K(t - \tau ) U_{\parallel}
( \tau )|\psi >_{||} d \tau,   \label{KR1}
\end{equation}
where $ U_{\parallel} (0)  =  P$,
\begin{equation}
K(t)  =  {\mit \Theta} (t) PHQ \exp (-itQHQ)QHP, \label{K}
\end{equation}
and ${\mit \Theta} (t)  =  { \{ } 1 \;{\rm for} \; t \geq 0, \; \;
0 \; {\rm for} \; t < 0 { \} }$.

The integro--differential equation (\ref{KR1}) can be replaced by
the following differential one (see \cite{bull} --- \cite{7})
\begin{equation}
( i \frac{\partial}{ {\partial} t} - PHP - V_{||}(t) )
U_{\parallel}(t)|\psi >_{||} = 0, \label{KR2}
\end{equation}
where
\begin{equation}
PHP + V_{||} (t) \stackrel{\rm def}{=} H_{||}(t). \label{H||=def}
\end{equation}
Taking into account (\ref{KR1}) and (\ref{KR2}) or (\ref{l1}) one
finds from (\ref{l1a}) and (\ref{KR1})
\begin{equation}
V_{\parallel} (t) U_{\parallel} (t) = - i \int_{0}^{\infty} K(t -
\tau ) U_{\parallel} ( \tau ) d \tau \stackrel{\rm def}{=} - iK
\ast U_{\parallel} (t) . \label{V||=def}
\end{equation}
(Here the asterisk, $\ast$, denotes the convolution: $f  \ast g(t)
= \int_{0}^{\infty}\, f(t - \tau ) g( \tau  ) \, d \tau$
).\linebreak Next, using this relation and a retarded Green's
operator  $G(t)$ for the equation (\ref{KR1})
\begin{equation}
G(t) = - i {\mit \Theta} (t) \exp (-itPHP)P, \label{G}
\end{equation}
one obtains \cite{9,10}
\begin{equation}
U_{\parallel}(t) = \Big[ {\it 1} + \sum_{n = 1}^{\infty} (-i)^{n}L
\ast \ldots \ast L \Big] \ast U_{\parallel}^{(0)} (t) ,
\label{U||-szer}
\end{equation}
where $L$ is convoluted $n$ times, ${\it 1} \equiv {\it 1}(t)
\equiv \delta (t)$,
\begin{equation}
L(t) = G \ast K(t), \label{L=GK}
\end{equation}
\begin{equation}
U_{\parallel}^{(0)} = \exp (-itPHP) \; P \label{U0}
\end{equation}
is a "free" solution of Eq. (\ref{KR1}). Thus from (\ref{V||=def})
\begin{equation}
V_{\parallel}(t) \; U_{\parallel}(t) = - i K \ast \Big[ {\it 1} +
\sum_{n = 1}^{\infty} (-i)^{n}L \ast \ldots \ast L \Big] \ast
U_{\parallel}^{(0)} (t) , \label{V-szer}
\end{equation}
Of course, the  series (\ref{U||-szer}), (\ref{V-szer}) are
convergent if $\parallel L(t)
\parallel < 1$. If for every $t \geq 0$
\begin{equation}
\parallel L(t) \parallel \ll 1, \label{L<1}
\end{equation}
then, to the lowest order of  $L(t)$,  one  finds  from
(\ref{V-szer}) \cite{9,10}
\begin{equation}
V_{\parallel}(t) \cong V_{\parallel}^{(1)} (t) \stackrel{\rm
def}{=} -i \int_{0}^{\infty} K(t - \tau ) \exp {[} i ( t - \tau )
PHP {]} d \tau . \label{V||=approx}
\end{equation}
Thus \cite{8,pra,9,10}
\begin{equation}
H_{\parallel}(0) \equiv PHP, \; \; V_{\parallel}(0) = 0, \; \;
V_{\parallel} (t \rightarrow 0) \simeq -itPHQHP. \label{H||(0)}
\end{equation}

If the  projector $P$ is defined as in (\ref{P}) and $H$ has the
following property
\begin{equation}
PHP \equiv m_{0}\, P, \label{P-H12=0}
\end{equation}
that is for
\begin{equation}
H_{12} = H_{21} = 0,  \label{H12=0}
\end{equation}
the  approximate formula (\ref{V||=approx}) for $V_{\parallel}(t)$
leads to the following form of $P e^{itPHP}$,
\begin{equation}
P e^{\textstyle{ i t PHP}} = P e^{\textstyle{itm_{0}}},
\label{exp-PHP-H}
\end{equation}
and thus to
\begin{equation}
V_{\parallel}^{(1)} (t) = - PHQ \frac{e^{\textstyle{-it(QHQ -
m_{0})}} - 1}{QHQ - m_{0}} QHP, \label{V||(t)-H0}
\end{equation}
which leads to $V_{||} \stackrel{\rm def}{=} \lim_{t \rightarrow
\infty} V_{||}^{(1)} (t)$,
\begin{equation}
V_{||} = - \Sigma (m_{0}). \label{V-H-0}
\end{equation}
This means that in the case (\ref{P-H12=0})
\begin{equation}
H_{||} = m_{0} \, P - \,\Sigma (m_{0}), \label{H||-H12=0}
\end{equation}
and  $H_{||} = H_{LOY}$.

On the other hand, in the case
\begin{equation}
H_{12} = H_{21}^{\ast} \neq 0, \label{H12n0}
\end{equation}
the form of $Pe^{itPHP}$ is more complicated. For example in the
case of conserved CPT,  formula (\ref{V||=approx})  leads to the
following form of $V_{||} \stackrel{\rm def}{=} \lim_{t
\rightarrow \infty} V_{||}^{(1)} (t)$ \cite{improved,Piskorski}
\begin{eqnarray}
V_{||}^{\mit\Theta} & = & - \frac{1}{2} \Sigma (H_{0} +
|H_{12}|)\, \Big[ \Big( 1 - \frac{H_{0}}{|H{_{12}|}} \Big)P +
\frac{1}{|H_{12}|} PHP \Big]
\nonumber \\
& &  - \frac{1}{2} \Sigma (H_{0} - |H_{12}|)\, \Big[ \Big( 1 +
\frac{H_{0}}{|H{_{12}|}} \Big)P - \frac{1}{|H_{12}|} PHP \Big],
\label{V-H12n0}
\end{eqnarray}
where
\begin{equation}
H_{0} \stackrel{\rm def}{=} \frac{1}{2} ( H_{11} + H_{22} ),
\label{H0}
\end{equation}
and $V_{\parallel}^{\mit \Theta}$ denotes $V_{\parallel}$ when
(\ref{[CPT,H]=0}) occurs.

In the general case (\ref{H12n0}), when there are no assumptions
on symmetries of the type CP--,  T--, or CPT--symmetry for the
total Hamiltonian  H of the system considered, the form of $V_{||}
= V_{\parallel}(t \rightarrow \infty ) \cong V_{\parallel}^{(1)} (
\infty )$ is even more complicated. In such a case one finds the
following expressions for the matrix elements  $v_{jk}(t
\rightarrow \infty ) \stackrel{\rm def}{=} v_{jk}$  of
$V_{\parallel}$ \cite{9,10},
\begin{eqnarray}
v_{j1} = & - & \frac{1}{2} \Big( 1 + \frac{H_{z}}{\kappa} \Big)
{\Sigma}_{j1} (H_{0} + \kappa ) - \frac{1}{2} \Big( 1 -
\frac{H_{z}}{\kappa} \Big)
{\Sigma}_{j1} (H_{0} - \kappa )\nonumber   \\
& - & \frac{H_{21}}{2 \kappa} {\Sigma}_{j2} (H_{0} + \kappa )
+ \frac{H_{21}}{2 \kappa} {\Sigma}_{j2} (H_{0} - \kappa ) ,
\nonumber \\
&  & \label{v-jk}\\
v_{j2} = & - & \frac{1}{2} \Big( 1 - \frac{H_{z}}{\kappa} \Big)
{\Sigma}_{j2} (H_{0} + \kappa ) - \frac{1}{2} \Big( 1 +
\frac{H_{z}}{\kappa} \Big)
{\Sigma}_{j2} (H_{0} - \kappa ) \nonumber  \\
& - & \frac{H_{12}}{2 \kappa} {\Sigma}_{j1} (H_{0} + \kappa ) +
\frac{H_{12}}{2 \kappa} {\Sigma}_{j1} (H_{0} - \kappa ) ,\nonumber
\end{eqnarray}
where $j,k = 1,2$,
\begin{equation}
H_{z} = \frac{1}{2} ( H_{11} - H_{22} ) , \label{H-z}
\end{equation}
and
\begin{equation}
\kappa = ( |H_{12} |^{2} + H_{z}^{2} )^{1/2} . \label{kappa}
\end{equation}
Hence, by (\ref{H||=def})
\begin{equation}
h_{jk} = H_{jk} + v_{jk} . \label{KR-h-jk}
\end{equation}
It should be emphasized that  all components  of the expressions
(\ref{v-jk}) are of the same order with respect  to $\Sigma (
\varepsilon )$.

In the case of preserved  CPT--symmetry  (\ref{[CPT,H]=0}),  one
finds $H_{11} = H_{22}$ which implies that $\kappa \equiv |H_{12}
|$, $H_{z} \equiv 0$ and  $H_{0} \equiv H_{11} \equiv H_{22}$, and
\cite{9,10}
\begin{equation}
{\Sigma}_{11} ( \varepsilon = {\varepsilon}^{\ast} ) \equiv
{\Sigma}_{22} ( \varepsilon = {\varepsilon}^{\ast} ) \stackrel{\rm
def}{=} {\Sigma}_{0} ( \varepsilon = {\varepsilon}^{\ast} ) .
\label{sigma-11}
\end{equation}
Therefore  matrix  elements  $v_{jk}^{\mit \Theta}$  of operator
$V_{\parallel}^{\mit \Theta}$  take the following form
\begin{eqnarray}
v_{j1}^{\mit \Theta} = & - & \frac{1}{2} {\Big\{ } {\Sigma}_{j1}
(H_{0} + | H_{12} |)
+ {\Sigma}_{j1} (H_{0} - | H_{12} |)  \nonumber \\
& + & \frac{H_{21}}{|H_{12}|} {\Sigma}_{j2} (H_{0} + | H_{12} |) -
\frac{H_{21}}{|H_{12}|} {\Sigma}_{j2} (H_{0} - | H_{12} |) {\Big\}
} , \nonumber \\
& & \label{v-jk-cpt} \\
v_{j2}^{\mit \Theta} = & - & \frac{1}{2} {\Big\{ } {\Sigma}_{j2}
(H_{0} + | H_{12} |)
+ {\Sigma}_{j2} (H_{0} - | H_{12} |)  \nonumber \\
& + & \frac{H_{12}}{|H_{12}|} {\Sigma}_{j1} (H_{0} + | H_{12} |) -
\frac{H_{12}}{|H_{12}|} {\Sigma}_{j1} (H_{0} - | H_{12} |) {\Big\}
}, \nonumber
\end{eqnarray}

Assuming
\begin{equation}
|H_{12}| \ll |H_{0}|, \label{H<H0}
\end{equation}
we find
\begin{equation}
v_{j1}^{\mit  \Theta} \simeq - {\Sigma}_{j1} (H_{0} ) - H_{21}
\frac{ \partial {\Sigma}_{j2} (x) }{\partial x}
\begin{array}[t]{l} \vline \, \\ \vline \,
{\scriptstyle x = H_{0} } \end{array} , \label{v-j1-cpt<}
\end{equation}
\begin{equation}
v_{j2}^{\mit  \Theta} \simeq - {\Sigma}_{j2} (H_{0} ) - H_{12}
\frac{ \partial {\Sigma}_{j1} (x) }{\partial x}
\begin{array}[t]{l} \vline \, \\ \vline \,
{\scriptstyle x = H_{0} } \end{array} , \label{v-j2-cpt<}
\end{equation}
where $j = 1,2$.  One  should  stress  that  due  to  the presence
of resonance terms, derivatives $\frac{\partial}{\partial x}
{\Sigma}_{jk} (x)$ need not  be  small and the same is true about
products $H_{jk} \frac{\partial}{\partial x} {\Sigma}_{jk}  (x)$
in  (\ref{v-j1-cpt<}), (\ref{v-j2-cpt<}). Finally, assuming that
(\ref{H<H0}) holds and using relations (\ref{v-j1-cpt<}),
(\ref{v-j2-cpt<}), (\ref{KR-h-jk}) and the expression (\ref{b5}),
we obtain for the CPT--invariant system \cite{is,hep-ph-0202253}
\begin{equation}
h_{j1}^{\mit  \Theta} \simeq  h_{j1}^{\rm LOY} - H_{21} \frac{
\partial {\Sigma}_{j2} (x) }{\partial x}
\begin{array}[t]{l} \vline \, \\ \vline \,
{\scriptstyle x = H_{0} } \end{array}  \stackrel{\rm def}{=}
h_{j1}^{\rm LOY} + \delta h_{j1}, \label{h-j1-cpt<}
\end{equation}
\begin{equation}
h_{j2}^{\mit  \Theta} \simeq  h_{j2}^{\rm LOY} - H_{12} \frac{
\partial {\Sigma}_{j1} (x) }{\partial x}
\begin{array}[t]{l}  \vline \, \\ \vline \,
{\scriptstyle x = H_{0} } \end{array}  \stackrel{\rm def}{=}
h_{j2}^{\rm LOY} + \delta h_{j2}, \label{h-j2-cpt<}
\end{equation}
where $j = 1,2$. From these formulae we  conclude  that, e.g., the
difference  between the diagonal   matrix   elements   of
$H_{\parallel}^{\mit \Theta}$ which plays  an  important  role in
designing CPT--invariance tests  for  the  neutral  kaons  system,
equals

\begin{equation}
\Delta h \stackrel{\rm def}{=} h_{11} - h_{22} \simeq H_{12}
\frac{ \partial {\Sigma}_{21} (x) }{\partial x}
\begin{array}[t]{l} \vline \, \\ \vline \,
{\scriptstyle x = H_{0} } \end{array} - H_{21} \frac{ \partial
{\Sigma}_{12} (x) }{\partial x}
\begin{array}[t]{l} \vline \, \\ \vline \,
{\scriptstyle x = H_{0} }\end{array} \neq 0. \label{delta-h}
\end{equation}

\section{Final remarks}

In the case of conserved CPT-- and violated CP-- symmetries there
must be
$$ h_{11}^{\mit \Theta}(t) - h_{22}^{\mit \Theta}(t) \neq 0 \; \;
{\rm for} \; \; t > t_{0} =0, $$ and, $ h_{11}(0) = h_{22}(0) =
<{\bf 1}|H|{\bf 1}>$, for the exact $H_{||}$.

Note that properties of the more accurate approximation described
in Sec. 4 are consistent with the general properties and
conclusions obtained in Sec. 3 for the exact effective Hamiltonian
--- compare (\ref{H||(0)}) and (\ref{hjjt=0}) and relations
(\ref{h11-h22=0}) with (\ref{delta-h}).

From the result ( \ref{delta-h}) it follows that $\Delta h = 0$
can be achieved only if $H_{12}=H_{21} = 0$. This means that if
the first order $|\Delta S| = 2$ interactions are forbidden in the
$K_{0}, {\overline{K}}_{0}$ complex then predictions following
from the use of the mentioned more accurate approximation  and
from the LOY theory should lead to the the same masses for $K_{0}$
and for ${\overline{K}}_{0}$. This does not contradict the results
of Sec. 3 derived for the exact $H_{||}$: the mass difference is
very, very  small and should arise at higher orders of the more
accurate approximation.

On the other hand from (\ref{delta-h}) it follows that $\Delta h
\neq 0$ if and only if $H_{12} \neq 0$. This means that if
measurable deviations from the LOY predictions concerning the
masses  of, e.g. $K_{0}, {\overline{K}}_{0}$ mesons are ever
detected, then the most plausible interpretation of this result
will be the existence of first order $|\Delta S| = 2$ interactions
in the system considered.

\end{document}